\documentstyle[12pt]{article}
\newcommand{\Omb}{\Omega _b}
\newcommand{\Omc}{\Omega _c}
\newcommand{\h}{\hat}
\begin{document}

\begin{titlepage}
\begin{center}
\large
\bf
 {$ \Omb \rightarrow \Omc , \Omc ^* $  transitions:\\
    Model-independent bounds on invariant form factors.}\\[2cm]
\rm
J.G.K\"{o}rner$^1$, K.Melnikov$^{2*}$, O.Yakovlev$^{3**}$\\[.5cm]
Johannes Gutenberg-Universit\"at\\
Institut f\"ur Physik (THEP), Staudingerweg 7\\
D-55099 Mainz, Germany\\
\normalsize
\bf
Abstract\\
\rm
\small
\end{center}
 In this note we report some model-independent bounds
involving transition form factors for $ \Omb \rightarrow \Omc $
and $ \Omb \rightarrow \Omc ^* $ and the nonperturbative matrix elements of
the $ \Omb $ system. They are derived by using
operator product expansion (OPE) in Heavy Quark Effective Theory .

[5cm]
\footnotesize
$^1$ Supported in part by the BMFT, Germany, under contract 06MZ730 \\
$^2\,$Supported by the Graduiertenkolleg Teilchenphysik, Universit\"at Mainz\\
$^3$ Supported by Deutsche Forshcungsgemeinschaft\\
$^{*)}$ on leave of absence from
 Novosibirsk State University, Novosibirsk, Russia \\
$^{**)}$ on leave of absence from
Budker Institute for Nuclear Physics, Novosibirsk, Russia \\[2cm]

\normalsize
\end{titlepage}

1. Recently it has been shown how to obtain  model-independent bounds
on the formfactors that describe semileptonic decays of heavy hadrons.
In what follows, we extend previous results on mesonic transitions
$ B \rightarrow D, D^* $ [1-3] and the baryonic transitions
$ \Lambda _b \rightarrow \Lambda _c $ [4] to the baryonic transitions
 $ \Omb \rightarrow \Omc ,
 \Omc ^* $.

2. As usual, we start from the time-ordered product of two appropriate currents
(vector or axial) inducing $ b \rightarrow c $ transitions.
\begin{equation}T_{\mu \nu}= -i\int \mbox {d}^4\! xe^{-iq\cdot x}
{J^+}_\mu (x) J_\nu (0)
\end{equation}

The best way to obtain the appropriate
operator product expansion is to integrate out
 the intermediate quark field which  reduces Eq.(1) to:

\begin{equation}T_{\mu \nu}= -i\int\mbox {d}^4\! xe^{-iq\cdot x} \langle x |
\bar Q \Gamma  \frac {1}{i\h D-m_c+i\epsilon} \Gamma Q |0 \rangle
\end{equation}

Here $ \langle x | $  and $ | 0 \rangle $
stand for states with definite space coordinates.
The usual HQET fermion field redifinition $ Q = e^{im_b v \cdot X} h $
results in the simple replacement of $ iD \rightarrow m_bv + iD $ in
Eq.(2) . (Here X stands for the four-space coordinate operator).
Then rewriting $ e^{-iq\cdot x} \langle x | $ as
$ \langle x | e^{-iq \cdot X} $ we finally get the
 well-known formulae which is
at the basis of the operator product expansion method in HQET [11]:
\begin{equation}
T_{\mu \nu}= -i\int\mbox{d}^4 x \langle x |
\bar Q \Gamma  \frac {1}{m_b\h v -\h q+i\h D-m_c+i\epsilon} \Gamma Q |0 \rangle
\end{equation}

As is seen from the redefinition of the quark field, the covariant
derivative in Eq.(3) correspondes to the "residual" momenta of the
heavy quark inside the heavy hadron. The natural assumption for this
quantity is to be of $ O( \Lambda_{QCD} ) $ . Assuming also $ q \approx
m_b $ we recognize the possibility to expand Eq.(3) in powers of
$ ( \frac {iD}{m_b} ) $. In what follows we will expand Eq.(3) up to
second order in  $ ( \frac {iD}{m_b} ) $ and consider  matrix element of the
 time-ordered
product between
 $ \Omb $ states.

3. Expanding Eq.(3) up to second order in $ ( \frac {iD}{m_b} ) $
we obtain  terms with  two, one or no derivatives.
Let us generically write the corresponding matrix elements as

\begin{eqnarray}
&&{\langle \Omc | \bar h \Gamma _1 h | \Omc \rangle }\\ &&
\langle \Omc | \bar h \Gamma _1 iD_{\alpha } h | \Omc \rangle \\ &&
\langle \Omc | \bar h \Gamma _1 iD_{\alpha } iD{\beta} h
| \Omc \rangle \
\end{eqnarray}

where $ \Gamma _1 $ stands for an arbitrary $ 4*4 $  matrix in spinor space.
\par
Now we are in the position to discuss  how to compute these matrix elements.
To begin with, let us discuss the matrix element ( Eq.(6) )
 with two derivatives.
As we are interested in the power corrections of  order
$ (\frac {\Lambda _{QCD}}{m_b} )^2 $  we need this matrix element only
up to the zeroth order in this parameter. It is then straightforward
to eliminate  residual $ h $ fields in favour of the HQET
fields $ h_v $  and to use the spin-flavour symmetry together with the
 lowest order HQET
equations of motion to parametrize this matrix element. We get:
\begin{eqnarray}
 \frac { \langle \Omb | \bar h \Gamma _1 iD_{\alpha } iD_{\beta } h
| \Omb \rangle }{{m_b}^2} & = &
u_1\cdot (g_{\alpha \beta }-v_\alpha v_\beta ) \bar R^a \Gamma _1
R_a \nonumber \\ + u_2 \cdot ( \bar R_\alpha \Gamma _1 R_\beta +
\bar R_\beta \Gamma _1 R_\alpha )
  &+& u_3 \cdot (\bar R_\alpha \Gamma _1 R_\beta -
\bar R_\beta \Gamma _1 R_\alpha )
\end{eqnarray}

Here $ R^a $ stands for the $ \Omb $ state in HQET (see Refs. [5-9] ).
The constants  $ u_1 $ , $ u_2 $  ,$  u_3 $, appearing in Eq.(7)
 are of order $(\frac {\Lambda
_{QCD}}{m_b})^2
 $
 Note that in contrast to the previously treated  meson and baryon
transitions,
here we need one more invariant matrix element
to parametrize Eq.(7).
The reason is that the total angular momenta of
the light diquark system in the hadron is equal to one vs.
$ J_{light}= \frac {1}{2} $ and $ J_{light}=0 $ in the previously
 studied cases.
 Taking for example $ \Gamma_1 $ to be the unity
matrix and contracting Eq.(7) with the metric tensor, we obtain:
\begin{equation}
\frac { \langle \Omb | \bar h  {(iD)}^2 h
| \Omb \rangle }{{m_b}^2} = - ( 3u_1+2u_2)
\end{equation}
Using the normalization $ \bar R^a R_a = -1 $  the quantity on the left hand
 side of Eq.(8) represents the heavy
quark kinetic energy inside the $ \Omb $  baryon. A similar quantity
 was introduced earlier for mesons in  [1-3] and estimated recently in [11].
Here we use similar notation and define:

\begin{equation}
\frac { \langle \Omb | \bar h  {(iD)}^2 h
| \Omb \rangle}{{m_b}^2} = - ( 3u_1+2u_2)= \frac {-{\mu_ {\pi}}^2}
{{m_b}^2}
\end{equation}

Analogiously, setting $ \Gamma _1 $ equal to $ 2 i {\sigma}
 _{ \alpha \beta } $ we arrive at:

\begin{equation}
\frac {- { \langle \Omb | \bar h {\sigma}
 _{ \alpha \beta } g_{QCD} G^{ \alpha \beta }  h
| \Omb \rangle}}{{m_b}^2} = 2u_3 R^{\beta } {\sigma}
 _{ \alpha \beta } R^{\alpha}=8 u_3= 2 \frac {{\mu _G}^2}{m^2}
\end{equation}

Here we have introduced an additional quantity $ {\mu _G }^2 $ , which can be
related to the mass difference of the $   \Omb $  and $ \Omb ^* $  baryons:
\begin{equation}
 m_{ \Omb ^*} - m_{\Omb }= \frac {3 {\mu _G}^2}{8 m_b}
\end{equation}
\par
Next let as discuss the calculation of the matrix element  Eq.
(5)
containing
one derivative. This matrix element is only of order  $ {\Lambda _{QCD}}
\over m_b $  , hence we have to calculate it in the next to leading order
in this parameter. First we expand the QCD fields up to the first order in
$ {\Lambda _{QCD}}
\over m_b $:

\begin{equation}
 h(x)=(1+\frac {i\h D}{2m} ) h_v(x)
\end{equation}

We then arrive at the following expression:
\begin{eqnarray}
 \langle \Omb | \bar h \Gamma _1 iD_{\alpha } h | \Omb \rangle
&=&
 \langle \Omb | \bar h_v  \Gamma _1 i D_\alpha h_v |
\Omb \rangle + \nonumber \\
\frac {1}{2m_b} \langle \Omb | \bar h_v (-i\h D \Gamma _1 iD_{\alpha}
&+& \Gamma _1 iD_{\alpha} i\h D ) h_v |\Omb \rangle
\end{eqnarray}
However, it can be shown that the first term on the r.h.s. of Eq.(13)
is  of  order $ (\frac {\Lambda _{QCD}}{m_b})^2 $ .
 The proof is based on the equation
 of motion of the heavy quark and the observation that
 the first term on the r.h.s.
is proportional to the velocity of the heavy quark up to the required accuracy.
In this way we finally obtain:
\begin{eqnarray}
&& \langle \Omb | \bar h \Gamma _1 iD_{\alpha} h | \Omb \rangle
=  \frac {(-v_\alpha \langle \Omb | \bar h_v \Gamma _1 {(i\h D_{\perp})}^2 h_v
|
\Omb \rangle }{2m_b} \nonumber \\  \nonumber \\
&+& \frac {\langle \Omb | \bar h_v (-i\h D \Gamma _1
 iD_{\alpha}
+ \Gamma _1 iD_{\alpha} i\h D ) h_v |\Omb \rangle }{2m_b}
\end{eqnarray}

This last expression has to be computed
up to the leading order in $ {\Lambda _{QCD}}
\over m_b $, \\ consequently
we can use the Eq.(7) to rewrite  the result in terms of the parameters $ u_1
$,
$ u_2 $ and $  u_3 $.

Let us finally consider the matrix element Eq.(4) with no derivatives
$ \langle \Omb | \bar h \Gamma _1 h | \Omb \rangle $.
In order to obtain this matrix element up to the required  order
one has to expand $ \bar h \Gamma _1 h $ up to the second order
in the inverse powers of the quark mass and one also has to take into account
the difference between the QCD and HQET wave functions of the final
and initial states. In this way we obtain the following parametrization
for the full basis of $ 4*4 $ matrices:

\begin{eqnarray}
&& \langle \Omb | \bar h h | \Omb \rangle = 1-\frac {{\mu _\pi}^2}
{2{m_b}^2}+ \frac {{\mu _G}^2}{2{m_b}^2} \\
&& \langle \Omb | \bar h \gamma _\mu h | \Omb \rangle = v_\mu \\
&& \langle \Omb | \bar h \gamma _\mu \gamma _5 h | \Omb \rangle
= \frac {-s _\mu}{3} ( 1+ \frac {{\mu _s}^2}{{m_b}^2}) \\ &&
\langle \Omb | \bar h \sigma _{ \mu \nu } h | \Omb \rangle = \frac
{i \epsilon _{\mu \nu \alpha \beta } v _\alpha s_\beta }{3}
(1 + \frac {{\mu_s}^2}{{m_b}^2} + \frac{u_1 + 14 u_2-4 u_3}{2})\\ &&
\langle \Omb | \bar h \gamma _5 h | \Omb \rangle =0
\end{eqnarray}

4. Using the parametrization of the matrix elements discussed in the
previous section it is straightforward to compute
$ \langle \Omb | T_{\mu \nu } | \Omb \rangle $  up to the necessary order
after
expanding equation (3) in terms of $ {iD} \over m $ . The result for the
invariant form factors (which are defined in full analogy with refs. [2,3,4])
are too lenghty to be presented here. What is really of interest is the
zero-recoil projection of these quantaties onto the helicity structure
functions.
 We thus compute  $ {n_\mu }^{*(\lambda )}{n_\nu }^{\lambda } \int d( q \cdot
v ) Im \langle \Omb | T_{\mu \nu } | \Omb \rangle $ at the point of zero recoil
. Here $ n^{(\lambda )} $ stands for the set of polarization vectors of the
outgoing  particle
with helicity $ \lambda $ ( say, $ W $ - boson for the weak-current case ).
This quantity is positive definite since it is in one-to-one correspondence
with the particle decay width into the diagonal helicity states of the
, off-
shell $ W $'s.
\par
5. On the other hand, we can express $ T_{\mu \nu } $ in terms of the
phenomenological form factors which describe the $ \Omb \rightarrow \Omc $,
$ \Omb \rightarrow \Omc ^* $ and $ \Omb \rightarrow $ excited states
transitions. Again, projecting  "hadronic" tensor $ T_{\mu \nu } $
onto helicity states, taking the imaginary part and integrating over
$ q \cdot v $ we arrive at the positive definite quantities $ W_L $,
$W_{T_{L,R}}$ and $ W_0 $. ( for further details see ref. [4] ).
The exact expressions of $ \Omc $ and $\Omc ^* $ contributions
to this quantities
for the case of axial and vector current are presented in the Appendix.
\par
6. As a next point let as discuss the sum rules for the form factors.
\par
Taking linear combination
$ \frac {1}{2} ( W_{T_L} + W_{T_R} ) $ for the hadron-side and parton-side
contributions  and neglecting the contributions from the excited
states, we finally get the inequality:
\begin{equation}
{|{f_1}^A|}^2 + \frac {2}{3} {|{G_1}^A|}^2 \leq
1-\frac {{\mu _\pi }^2}{{m_b}^2}(\frac {x^2}{4} + \frac {x}{6} +\frac
{1}{4} )-
\frac {{\mu_G}^2}{{m_b}^2}(\frac {x^2}{12} - \frac {x}{6} -\frac {1}{4} )
\end{equation}
Here $ x $ stands for the ratio  $ \frac {m_b}{m_c} $.
\par
On the other hand, taking $ n \cdot s = -1 $ and calculating $ W_{T_L} $
we can " switch off " the $ \Omb $ state contribution on the hadron side,
thus obtainig a
sum rule  for the $ {G_1}^A $ formfactor only:
\begin{equation}
\frac {1}{2} {|{G_1}^A|}^2 \leq \frac {2}{3} - \frac {{\mu _s}^2}{3{m_b}^2}
+ \frac {{\mu _G }^2}{{m_b}^2} ( \frac {1}{3}+ \frac {x}{6} )-
\frac {{\mu _\pi}^2}{{m_b}^2}(\frac {x^2}{6}+\frac {x}{6} +\frac {1}{4})+
u_1(\frac {x}{6}- \frac {1}{12} ) - u_2 (x+\frac {14}{12} )
\end{equation}
Note that $ u_1$ and $ u_2 $ enter this sum rule as independent
quantities i.e. they do not enter in the combination of kinetic energy (7).
\par
In full analogy  we can obtain a bound for the vector form factors.
 It is worth noting, that at zero recoil, the  $ \Omc ^* $ state does not
contribute to the vector-current induced transition (see Appendix).
 Thus we get:
\begin{equation}
{|\Sigma {f_i}^V|}^2 \leq 1- \frac {( {\mu _\pi }^2-{\mu _G}^2)}{4{m_b}^2}
{(x-1)}^2
\end{equation}
As $ \Omc ^* $ does not contribute to the vector current induced transitions,
we can also think of "switching off " the contribution from the $ \Omc $
state. We can do that by taking the first moment while integrating over $
q \cdot v $ ( similar to the   Voloshin  sum rule , but at zero recoil ).
The leading contributions to the " hadronic " side of the sum rules will be
zero in this way , while we get something none zero on the partonic side.
Again neglecting the  contributions from the excited states  we get:
\begin{equation}
{\mu _G}^2 \leq {\mu _\pi }^2
\end{equation}
We mention here that a similar inequality was obtained in the Ref.[11]
 for mesonic states using different
techniques.
\par
7. To summarize, we have estimated
the size of $ (\frac {\Lambda _{QCD}}{m_b})^2
$ corrections to  $ \Omb $ to $ \Omc $ transitions at zero recoil point
using the operator product expansion in HQET.
\par
8. {\bf Acknowledgments. } The authors are grateful to  Dan Pirjol
for useful conversations.

\begin{center}  \bf { Appendix. }
\end{center}

The contribution of $\Omc$ , $\Omc ^* $ to the projection of the hadronic
tensor to the helicity structure functions:
\begin{eqnarray}
\Omb \to \Omc
\nonumber \end{eqnarray}
1.Axial current:
\begin{eqnarray}
<\Omc (v^{'},s^{'})\mid \bar c \gamma_{\mu}\gamma_{5}b\mid \Omb (v,s)>&=&
\bar u_c(v^{'},s^{'})[f_1^A\gamma_{\mu}+f_{2}^{A}v_{\mu}+
f_{3}^Av^{'}_{\mu} ]\gamma_{5}u_b(v,s);
\nonumber \end{eqnarray}
\begin{eqnarray}
W_L&=&\frac{w+1}{2q^2}((m_{\Omb}-m_{\Omc})f_1^A-
m_{\Omc}(w-1)f_2^A-m_{\Omb}(w-1)f_3^A)^2;
\nonumber \end{eqnarray}
\begin{eqnarray}
W_{T_{L,R}}=\mid f^A_1 \mid^2 \frac{w+1}{2}(1\pm \vec s \vec n);
\nonumber \end{eqnarray}
\begin{eqnarray}
W_0=\frac{w-1}{2q^2}((m_{\Omb}+m_{\Omc})f_1^A-
(m_{\Omb}-m_{\Omc}w)f_2^A-(m_{\Omb}w-m_{\Omc})f_3^A)^2;
\nonumber \end{eqnarray}

2.Vector current:
\begin{eqnarray}
<\Omc (v^{'},s^{'}) \mid \bar c \gamma_{\mu}b\mid \Omb (v,s)>&=&
\bar u_c(v^{'},s^{'})[f_1^V\gamma_{\mu}+f_{2}^{V}v_{\mu}+
f_{3}^Vv^{'}_{\mu} ]u_b(v,s);
\nonumber \end{eqnarray}
\begin{eqnarray}
W_L&=&\frac{w-1}{2q^2}((m_{\Omb}+m_{\Omc})f_1^V+
m_{\Omc}(w+1)f_2^V+m_{\Omb}(w+1)f_3^V)^2;
\nonumber \end{eqnarray}
\begin{eqnarray}
W_{T_{L,R}}=\mid f^V_1 \mid^2 \frac{w-1}{2}(1\pm \vec s \vec n);
\nonumber \end{eqnarray}
\begin{eqnarray}
W_0=\frac{w+1}{2q^2}((m_{\Omb}-m_{\Omc})f_1^V+
(m_{\Omb}-m_{\Omc}w)f_2^V+(m_{\Omb}w-m_{\Omc})f_3^V)^2;
\nonumber \end{eqnarray}

\begin{eqnarray}
\Omb \to \Omc ^*
\nonumber \end{eqnarray}
1.Axial current:
\begin{eqnarray}
<\Omc ^*  \mid \bar c \gamma_{\mu}\gamma_{5}b\mid \Omb >&=&
\bar u_c^{\nu}(v_2)[G_1^Ag_{\nu ,\mu}\gamma_{5}+G_{2}^{A}v_{1\nu}
\gamma_{\mu}\gamma_{5}\nonumber\\
&+&
G_{3}^Av_{1\nu}v_{1\mu}\gamma_{5}+
G_{4}^Av_{1\nu}v_{2\mu}\gamma_{5}]
\gamma_{5}u_b(v_1);
\nonumber \end{eqnarray}
\begin{eqnarray}
W_L&=&\frac{w+1}{3q^2}((m_{\Omb}w-m_{\Omc ^* })G_1^A+
(m_{\Omc ^* }+m_{\Omb})(w-1)G_2^A\nonumber \\
&+&m_{\Omc ^*}(w^2-1)G_3^A+
m_{\Omb}(w^2-1)G_4^A)^2;
\nonumber \end{eqnarray}
\begin{eqnarray}
W_0&=&\frac{(w-1)(w+1)^2}{3q^2}(m_{\Omb}G_1^A+
(m_{\Omb}-m_{\Omc1})G_2^A\nonumber \\
&+&(m_{\Omb}-m_{\Omc ^*}w)G_3^A+
(m_{\Omb}w-m_{\Omc  ^*})G_4^A)^2;
\nonumber \end{eqnarray}
\begin{eqnarray}
W_{T_{L,R}}&=&\frac{w+1}{4}(
\frac{1}{3}(G_1^A-2(w-1)G^A_2)^2+
(G_1^A)^2\nonumber \\
&\pm&\vec s \vec n (
\frac{1}{3}(G_1^A-2(w-1)G^A_2)^2-
(G_1^A)^2);
\nonumber \end{eqnarray}

2.Vector current:
\begin{eqnarray}
<\Omc ^*  \mid \bar c \gamma_{\mu}b\mid \Omb >&=&
\bar u_c^{\nu}(v_2)[G_1^Vg_{\nu ,\mu}+G_{2}^{V}v_{1\nu}
\gamma_{\mu}\nonumber\\
&+&
G_{3}^Vv_{1\nu}v_{1\mu}+
G_{4}^Vv_{1\nu}v_{2\mu}]\gamma_{5}
u_b(v_1);
\nonumber \end{eqnarray}
\begin{eqnarray}
W_L&=&\frac{w-1}{3q^2}((m_{\Omb}w-m_{\Omc ^* })G_1^V-
(m_{\Omb}-m_{\Omc ^* })(w+1)G_2^V\nonumber \\
&+&m_{\Omc ^* }(w^2-1)G_3^V+
m_{\Omb}(w^2-1)G_4^V)^2;
\nonumber \end{eqnarray}
\begin{eqnarray}
W_0&=&\frac{(w+1)(w-1)^2}{3q^2}(m_{\Omb}G_1^V-
(m_{\Omb}+m_{\Omc ^* })G_2^V\nonumber \\
&+&(m_{\Omb}-m_{\Omc ^* }w)G_3^V+
(m_{\Omb}w-m_{\Omc ^* })G_4^V)^2;
\nonumber \end{eqnarray}
\begin{eqnarray}
W_{T_{L,R}}&=&\frac{w-1}{4}(
\frac{1}{3}(G_1^V-2(w+1)G^V_2)^2+
(G_1^V)^2\nonumber \\
&\pm&\vec s \vec n (
\frac{1}{3}(G_1^V-2(w+1)G^V_2)^2-
(G_1^V)^2);
\nonumber
\end{eqnarray}

\newpage
\begin{center} \bf {References}
\end{center}


1. J.Chay, H.Georgi, B.Grinstein, Phys. Lett. B247 (1990) 399.\\
2. I.Bigi, M.Shifman, N.Uraltsev, A.Vainstein, Phys. Rev. Lett. 71,(1993) 496
.\\
```B.Blok, L.Koyrakh, M.Shifman, A.Vainstein, Phys. Rev. D49 (1994) 3356. \\
3. A.Manohar, M.Wise , Phys. Rev. D49 (1994) 1310. \\
4. J.G. K\"{o}rner, D.Pirjol, preprint MZ-TH/94-17, May 94. \\
5. H.Georgi, Nucl. Phys. B348 (1991) 293.\\
6. T.Mannel, W.Roberts, Z.Ryzak, Nucl. Phys. B355, (1991) 38.\\
7. C.G Boyd, D.E.Brahm Phys. Lett. B257 (1991) 393.\\
8. A.Falk, Nucl. Phys. B378 (1992) 80. \\
9. J.G. K\"{o}rner, D.Pirjol, M. Kr\"{a}mer, Preprint,
   DESY 94-095, MZ-THEP-94-08.\\
10. I. Bigi, A.Grozin, M. Shifman, N.Uraltsev, A.Vainstein,\\
{}~~~~Preprint TPI-MINN-94/25-T.\\
11. M.Shifman, N.Uraltsev, A.Vainstein TPI-MINN/13-T.\\

\end{document}